\documentstyle[12pt,aasms,epsfig]{article}

\def \etal {et al.\thinspace}

\received{ 2000}
\journalid{337}{15 January 1989}
\articleid{11}{14}

\slugcomment{Revised and resubmitted to Astrophys. J. {\it Suppl.}, 2000}

\begin{document}

\title{Electron-Ion Recombination Rate Coefficients and Photoionization 
Cross Sections for Astrophysically Abundant 
Elements IV. Relativistic calculations for C~IV and C~V for UV and X-ray
modeling}

\author{Sultana N. Nahar, Anil K. Pradhan}
\affil{Department of Astronomy, The Ohio State University, 
Columbus, OH 43210}
\author{Hong Lin Zhang}
\affil{Applied Theoretical \& Computational Physics Division}
\affil{MS F663, Los Alamos National Laboratory, Los Alamos, NM 87545}

\begin{abstract}

 The first complete set of unified cross sections and rate coefficients are 
calculated for photoionization and recombination of He- and Li-like ions 
using the relativistic Breit-Pauli R-matrix method. We present total,
unified (e~+~ion) recombination rate coefficients for
$(e~+~C~VI \longrightarrow C~V)$ and $(e~+~C~V \longrightarrow C~IV)$
including fine structure.
 Level-specific recombination rate coefficients up to the n = 10
levels are also
obtained for the first time; these differ considerably from the
approximate rates currently available. Applications to recombination-cascade
coefficients in X-ray spectral models of K$\alpha$ emission from 
the important He-like ions are pointed out. The overall uncertainty in
the total recombination rates should not exceed 10-20\%.
Ionization fractions for Carbon are 
recomputed in the coronal approximation using the new rates.
 The present (e~+~ion) recombination rate coefficients are compared with
several sets of available data, including previous LS coupling results,
and `experimentally derived' rate coefficients. The role of relativistic
fine structure, resolution of resonances, radiation damping, and
interference effects is discussed. Two general features of
recombination rates are noted: (i) the non-resonant (radiative
recombination) peak as $E,T
\rightarrow 0$, and the (ii) the high-T resonant (di-electronic
recombination) peak.

\end{abstract}


\section{INTRODUCTION}

 Electron-ion recombination with H- and He-like ions is of particular
interest in X-ray astronomy (Proc. X-ray Symposium, 2000). X-ray emission 
in the K$\alpha$ complex of He-like ions, such as C~V, from the 
$n = 2 \rightarrow 1$ transitions yields 
perhaps the most useful spectral diagnostics for
temperature, density, ionization balance, and abundances
in the plasma source (Gabriel 1972, Mewe and Schrijver 1978, Pradhan and Shull
1981, Pradhan 1985). Li-like C~IV is of considerable importance in 
UV emission spectra from active galactic nuclei and quasars 
(e.g. Laor \etal 1994), as well as absorption in
AGN (Crenshaw and Kraemer 1999). In addition, the C~IV and other Li-like 
ionization states
are valuable tracers of the plasma in the `hot interstellar medium'
(Spitzer 1990, Spitzer and Fitzpatick 1993, Martin and Bowyer 1990,
Bregman and Harrington 1986). The
primary sets of atomic data needed for accurate calculations of
ionization fractions are for photoionization and recombination.

 Theoretical models of spectral formation also require excitation cross
sections and transition probabilities. A considerable amount of atomic 
data is being computed for these atomic
processes under the Iron Project (IP; Hummer \etal 1993)
for electron impact excitation and
radiative transition probabilities 
for astrophysically abundant  elements
using the Breit-Pauli R-matrix (BPRM) method that includes relativistic
fine structure in intermediate coupling (Berrington \etal 1995). The
present work is an extension of the IP work to photoionization and
recombination.

 The ionization balance equations usually correspond to photoionization
equilibrium

\begin{equation}
  \int_{\nu_0}^{\infty} \frac{4 \pi J_{\nu}}{h\nu} N(X^{z})
\sigma_{PI}(\nu,X^{z}) d\nu = \sum_j N_e N(X^{z+1}) \alpha_R(X_j^{z};T),
\end{equation}

and collisional equilibrium

\begin{equation}
C_I(T,X^{z}) N_e N(X^{z}) = \sum_j N_e N(X^{z+1}) \alpha_R(X_j^{z};T),
\end{equation}

\noindent
where the $\sigma_{PI}$ are the photoionization cross sections, and 
the $\alpha_R(X_j^{z};T)$ are the total electron-ion recombination rate
coefficients of the recombined ion of charge $z$, $X_j^{z}$, to state j
at electron temperature T. The $C_I$ are the rate coefficients for electron
impact ionization that can be reliably obtained from  experimental
measurements (Bell \etal 1983). On the other hand, the (e~+~ion) 
recombination cross sections and rates are difficult to compute or measure.
However, several experimental measurements of electron-ion
recombination cross
sections using ion storage rings have been carried out in recent years
(e.g. Wolf \etal 1991, Kilgus \etal
1990,1993; Mannervik \etal 1997). The experimental cross sections exhibit
detailed resonance structures observed
at very high resolution in beam energy, and measure absolute cross
sections. Therefore they provide ideal tests
for theoretical methods, as well as the physical effects included in the
calculations.  Many of these experimental measurements have been for
recombination with H- and He-like C and O. 

Among the recent theoretical developments is a self-consistent method
for calculations for photoionization and (e~+~ion) recombination, as 
described in previous papers in this series.
An identical eigenfunction expansion for the ion is employed in
coupled channel calculations for both processes, thus ensuring
consistently accurate cross sections and rates in an ab initio manner.
The theoretical treatment of (e~+~ion) recombination subsumes both
the non-resonant recombination (i.e. radiative recombination, RR), and
the resonant recombination (i.e. di-electronic recombination, DR)
processes in a unified scheme. In addition to the total, unified
recombination rates, level-specific recombination rates and
photoionization cross sections are obtained for a large number of
atomic levels. The calculations are carried out in the close coupling 
approximation using the R-matrix
method. Although the calculations are computationally intensive, they
yield nearly all photoionization and recombination parameters needed for
astrophysical photoionization models with higher precision than hitherto
possible.

 Previous calculations of unified (e~+~ion) recombination cross
sections and rates, reported in the present series on photoionization
and recombination, were carried out in LS coupling (Nahar and Pradhan
1997, paper I; Nahar 1999). There were two reasons. First, the
calculations are extremely complex and involve both radiative
photoionization and collisional electron-ion scattering calculations;
the full intermediate coupling relativistic calculations are many times more 
computationally intensive than the LS coupling ones. Second, the effect 
of relativistic fine structure was expected to be small
for these light elements. 

For the highly charged H- and He-like ions,
however, subsequent calculations showed that results including the 
relativistic effects are signficantly more accurate
not only in terms of more detailed resonance structure, but also
to enable a full resolution of resonances necessary to include radiative
damping (Pradhan and Zhang 1997, Zhang \etal 1999,
and references therein). The relativistic Breit-Pauli R-matrix
(BPRM) method is now extended to calculate the total and level-specific
recombination rate coefficients in the self consistent unified manner.
In this paper we describe the first of a series of full-scale BPRM 
calculations of photoionization and photo-recombination, as inverse 
processes, and DR, to obtain total, unified (e~+~ion) recombination 
rates of He- and Li-like Carbon, C IV and C V.

\section{THEORY}

The extension of the close coupling method to electron-ion recombination
is described in earlier works (Nahar \& Pradhan 1994, 1995), together with the 
details of the unified treatment. Here we present a brief description of 
the theory relevant to the calculations
of electron recombination cross sections with H-like and He-like ions.
The calculations are carried out in the close coupling (CC) approximation 
employing the $R$-matrix method  in intermediate coupling
with the BP Hamiltonian. The target ion is represented by an 
$N$-electron system, and the total wavefunction expansion, 
$\Psi(E)$, of the ($N$+1) electron-ion system of symmetry $SL\pi$ or
$J\pi$ may be represented in terms of the target eigenfunctions as:

\begin{equation}
\Psi(E) = A \sum_{i} \chi_{i}\theta_{i} + \sum_{j} c_{j} \Phi_{j},
\end{equation}
 
\noindent
where $\chi_{i}$ is the target wavefunction in a specific state $SL\pi$
or level $J_i\pi_i$ and $\theta_{i}$ is the wavefunction for the 
($N$+1)-th electron in a channel labeled as 
$S_iL_i(J_i)\pi_ik_{i}^{2}\ell_i(\ J\pi))$; $k_{i}^{2}$
being its incident kinetic energy. $\Phi_j$'s are the correlation
functions of the ($N$+1)-electron system that account for short range
correlation and the orthogonality between the continuum and the bound 
orbitals. 

In the relativistic BPRM calculations the set of ${SL\pi}$
are recoupled to obtain (e + ion) levels  with total $J\pi$, followed by
diagonalisation of the (N+1)-electron Hamiltonian,
\begin{equation}
H^{BP}_{N+1}\mit\Psi = E\mit\Psi.
\end{equation}
The BP Hamiltonian is
\begin{equation}
H_{N+1}^{\rm BP}=H_{N+1}+H_{N+1}^{\rm mass} + H_{N+1}^{\rm Dar}
+ H_{N+1}^{\rm so},
\end{equation}
where $H_{N+1}$ is the nonrelativistic Hamiltonian,
\begin{equation}
H_{N+1} = \sum_{i=1}\sp{N+1}\left\{-\nabla_i\sp 2 - \frac{2Z}{r_i}
        + \sum_{j>i}\sp{N+1} \frac{2}{r_{ij}}\right\},
\end{equation}
and the additional terms are the one-body terms, the mass correction term,
the Darwin term and the spin-orbit term respectively. Spin-orbit
interaction, $H^{so}_{N+1}$, splits the LS terms into fine-structure
levels labeled by $J\pi$, where $J$ is the total angular momentum.

The positive and negative energy states (Eq. 4) define continuum or
bound (e~+~ion) states.

\begin{equation}
 \begin{array}{l} E = k^2 > 0  \longrightarrow
continuum~(scattering)~channels \\  E = - \frac{z^2}{\nu^2} < 0
\longrightarrow bound~states, \end{array}
\end{equation}
where $\nu$ is the effective quantum number relative to the core level.
If $E <$ 0 then all continuum channels are `closed' and the solutions
represent bound states.

The photoionization cross section can be obtained as
\begin{equation}
\sigma_{PI} = {1\over g}{4\pi^2\over 3c}\omega{\bf S},
\end{equation}
where $g$ is the statistical weight factor of the bound state, ${\bf S}$
is the dipole line strength,
\begin{equation}
{\bf S}~=~|<\Psi_B || {\bf D} || \Psi_F >|^2
\end{equation}
and ${\bf D}$ is the dipole operator (e.g. Seaton 1987).

For highly charged ions (such as the H- and the He-like) radiative transition
probabilities in the core ion are very large and may be of the same
order of magnitude as autoionization probabilities.  Autoionizing
resonances may then undergo significant radiative decay and the
photoionization process may be written as

\begin{equation}
 h\nu + X^{+} \rightarrow (X^+)^{**}  \rightarrow  \left\{
\begin{array}{c} (i) \ e + X^{++} \\ (ii) \  h\nu' + X^+ \end{array} \right. 
\end{equation}

Branch (ii) represents radiation damping of autoionizing resonances.
In the present work
this radiative damping effect is included for all near-threshold
resonances, up to $\nu \leq 10$, using a resonance fitting procedure 
(Sakimoto et al. 1990, Pradhan and Zhang 1997, Zhang \etal 1999).

Recombination of an incoming electron to the target ion may occur through
non-resonant, background continuum, usually referred to as 
radiative recombination (RR),

\begin{equation}
e + X^{++} \rightarrow  h\nu + X^+,
\end{equation}

\noindent
which is the inverse process of direct photoionization, or through the
two-step recombination process via autoionizing resonances, i.e. 
dielectronic recombination (DR):

\begin{equation}
e + X^{++} \rightarrow (X^+)^{**}  \rightarrow  \left\{
\begin{array}{c} (i) \ e + X^{++} \\ (ii) \  h\nu + X^+ \end{array}
\right. ,
\end{equation}

\noindent
where the incident electron is in a quasi-bound doubly-excited state 
which leads either to (i) autoionization, 
a radiation-less transition to a lower state  of the ion and the free 
electron, or to (ii) radiative stabilization to recombining ion states
predominantly via decay of the ion core (usually to the ground state)
and the bound electron. 

In the unified treatment the photoionization cross sections, 
$\sigma_{\rm PI}$, of a large number of low-$n$ bound states -- all 
possible states with $n \leq n_{\rm max}\sim 10$ -- are obtained in the 
close coupling (CC) approximation as in the Opacity Project (Seaton 1987). 
Coupled channel calculations for $\sigma_{\rm PI}$ include both 
the background and the resonance structures (due to the doubly excited 
autoionizing states) in the cross sections. The recombination cross section, 
$\sigma_{\rm RC}$, is related to $\sigma_{\rm PI}$, through detailed balance 
(Milne relation) as

\begin{equation}
\sigma_{\rm RC}(\epsilon) = 
{\alpha^2 \over 4} {g_i\over g_j}{(\epsilon + I)^2\over \epsilon}\sigma_{\rm PI}
\end{equation} 
in Rydberg units; $\alpha$ is the fine structure constant, $\epsilon$ is
the photoelectron energy, and $I$ is the ionization potential. In the
present work, it is assumed that the recombining ion is in the ground
state, and recombination can take place into the ground or any of the 
excited recombined (e+ion) states.  Recombination rate coefficients of 
individual levels are obtained by averaging the recombination cross 
sections over the Maxwellian electron distribution, $f(v)$, at a given 
temperature as
\begin{equation}
\alpha_{RC}(T) = \int_0^{\infty}{vf(v)\sigma_{RC}dv}.
\end{equation}
The contributions of these bound states 
to the total $\sigma_{\rm RC}$ are obtained by summing over the contributions 
from individual cross sections. $\sigma_{\rm RC}$ thus obtained from 
$\sigma_{\rm PI}$, including the radiatively damped autoionizing resonances
(Eq. (10)), corresponds to the total (DR+RR) unified 
recombination cross section.

The recombination cross section, $\sigma_{RC}$ in Megabarns (Mb), is 
related to the collision strength, $\Omega_{\rm RC}$, as

\begin{equation}
\sigma_{RC}(i\rightarrow j)({Mb}) = \pi \Omega_{RC}(i,j)/(g_ik_i^2)
(a_o^2/1.\times 10^{-18}) ,
\end{equation}
where $k_i^2$ is the incident electron energy in Rydberg. As $\sigma_{RC}$
diverges at zero-photoelectron energy, the total collision strength,
$\Omega$, is used in the recombination rate calculations.

Recombination into the high-$n$ states must also be included, i.e. 
$n_{\rm max} < n \leq \infty$, (Fig.~1 of Nahar \& Pradhan 1994). To 
each excited threshold $S_iL_i(J_i)\pi_i$ 
of the $N$-electron target ion, there corresponds an infinite series of 
($N$+1)-electron states, $S_iL_i(J_i)\pi_i\nu\ell$, 
to which recombination can occur, where $\nu$ is the effective 
quantum number. For these states
DR dominates the recombination process and the background recombination
is negligibly small. The contributions from these
states are added by calculating the collision strengths, $\Omega_{\rm DR}$, 
employing the precise theory of radiation damping by Bell and Seaton 
(1985, Nahar \& Pradhan 1994). Several aspects related to the 
application of the theory to the calculation of DR collision strengths 
are described in the references cited. 
 General details of the theory and close coupling BPRM calculations are described in paper I and Zhang \etal (1999, and references therein).

\section{COMPUTATIONS}

The electron-ion recombination calculations entail CC calculations 
for photoionization and electron-ion scattering. Identical 
eigenfunction expansion for the target (core) ion is employed for 
both processes; thus enabling inherently self-consistent 
photoionization/recombination results in an ab initio manner for a
given ion. The total recombination cross sections, $\sigma_{\rm RC}$, 
are obtained from the photoionization cross sections, 
$ \sigma_{\rm PI}$, and DR collision strengths, $\Omega_{\rm DR}$, are 
calculated as descriped in paper I and Zhang \etal (1999).
However, the computations for the cross sections are repeated with 
a much finer energy mesh in order to delineate the detailed resonance 
structures as observed in the experiments. 

Computations of photoionization cross sections, $\sigma_{\rm PI}$, in
the relativistic BPRM intermediate coupling approximations are 
carried out using the package of codes from the Iron Project 
(Berrington \etal 1995; Hummer \etal 1993). However, radiation damping
of resonances up to $n=10$ are included through use of the extended 
codes of STGF and STGBF (Pradhan \& Zhang 1997). The $R$-matrix 
calculations are carried out for each total angular momentum symmetry 
$J\pi$, corresponding to a set of fine structure target levels $J_t$. 

In the energy region from threshold up to about $\nu = \nu_{\rm max}$
= 10 ($\nu$ is the effective quantum number of the outer orbital
of the recombined ion bound state), detailed photorecombination cross
sections are calculated as is Eq.~(12).  The electrons in
this energy range recombine to a large number of final (e~+~ion)
states and recombination cross sections are computed for all coupled
symmetries and summed to obtain the total $\sigma_{\rm RC}$.  The 
number of these final recombined states in the BPRM case is larger, 
owing to more channels involving fine structure, than the $LS$ 
coupling case.

In the higher energy region, $\nu_{\rm max} < \nu < \infty$ 
below each threshold target level, where the
resonances are narrow and dense and the background is negligible, we
compute the detailed and the resonance averaged DR cross sections.
The DR collision 
strengths in BPRM are obtained using extensions of the $R$-matrix 
asymptotic region codes (paper I; Zhang and Pradhan 1997) respectively.
It is necessary to use extremely fine energy meshes in order to
delineate the resonance structures belonging to each $n$-complex. 

The level specific recombination rate coefficients are obtained
using a new computer program, BPRRC (Nahar and Pradhan 2000). 
The program extends the photoionization
cross sections at the high energy region, beyond the highest target 
threshold in the CC wavefunction expansion of the ion, by a tail from 
Kramers fit of $\sigma_{PI}(E) = \sigma_{PI}^o(E_o^3/E^3)$. The level
specific rates are obtained for energies going up to infinity. 
These rates include both non-resonant and resonant contributions up to
energies $z^2/\nu_{\rm max}^2$; Contributions from all autoionizing
resonances up to $\nu \leq \nu_{\rm max} \approx 10$ are included.

The program BPRRC sums up the level specific rates, which is added to
the contributions from the high-n DR, to obtain the total recombination
rates. As an additional check on the numerical calculations, 
the total recombination rate 
coefficients, $\alpha_R$, are also calculated from the total recombination
collision strength,
obtained from all the photoionization cross sections,
and the DR collision strengths.
The agreement between the two numerical
approaches is within a few percent.

The background contribution from
the high-n states, ($10 < n \leq \infty$), to the total recombination
is also included as the "top-up" part (Nahar 1996). This contribution is 
important at low temperatures, but is negligible at high temperatures.
The rapid rise in $\alpha_R$ toward
very low temperatures is due to low energy recombination
into the infinite number of these high-n states, at electron energies
not usually high enough for resonant excitations.

Below we describe the calculations individually for the ions
under consideration.

\subsection{e + C V $\longrightarrow$ C IV}

The fine structure levels of the target ion, C V, included in the 
wavefunction expansion for C IV are given in Table 1. 
The 13 fine structure levels of C V up to $3p$ correspond to
configurations $1s^2$, $1s2s$, $1s2p$, $1s3s$, $1s3p$ (correlation
configurations also involve the n = 4 orbitals).
Although
calculated energies are within a few percent of the observed ones, the
latter are used in the computations for more accurate positions of 
the resonances. The bound channel wavefunction, second term in
$\Psi$ in Eq. (3), contains configuration $3d^2$. Levels of 
angular momentum symmeties
$1/2 \leq J \leq 9/2$ are considered. With largest partial wave of 
the outer electron is $l=9$, these correspond
to $0\leq L \leq 7$ in doublet and quartet spin symmetries. The R-matrix basis 
set is represented by 30 continuum functions. It is necessary to
represent the wavefunction expansion in the inner region of the R-matrix
boundary with a relatively large number of terms in order to avoid some
numerical problems that result in slight oscillation in computed cross
sections.

\subsection{e + C VI $\longrightarrow$ C V}

The wavefunction expansion of C V is represented by 9 fine structure
levels (Table 1) of hydrogenic C VI from $1s$ to $3d$. Correlation
orbitals $4s$, $4p$, $4d$, and $4f$ are also included. 
The $SL\pi$ symmetries consist of $0 \leq L
\leq 7$ of singlet and triplet spin symmetries, for even and odd parities. 
All levels 
of C~V with total angular momentum symmetry $0 \leq J \leq 6$ are 
included.  The R-matrix basis set consist of 40 terms to reduce 
numerical instablities that might otherwise result in slight
oscillations in the cross sections.

\section{RESULTS AND DISCUSSION}

 Results for photoionization and recombination are presented below, followed
by a discussion of the physical features and effects.

\subsection{Photoionization}

The ground state
cross sections are needed for various astrophysical models, such
as in determination of ionization fraction in photoionization
equilibrium of plasma.
 Figs. 1 and 2 present the ground state photoionization cross section
for C~IV ($1s^2 \ 2s \ (^2S_{1/2})$) and C~V $(1s^2 \ (^1S_0))$. Plots (a,b)
in each figure show the total cross section (a), and the partial cross
section (b) into the ground level of the residual ion.
The total cross sections Fig. 1(a) and Fig. 2(a) show the K-shell
ionization jump at the n = 2 target levels, i.e. inner-shell photoionization 

$$ h\nu + C~IV (1s^2 \ 2s,\ 2p) \longrightarrow e \ + 1s2s \ (,2p) \ , $$ 
and
$$ h\nu + C~V (1s^2, \ 1s2p) \longrightarrow e \ + 2s \ (,2p) \ . $$

 In X-ray photoionization models these inner-shell edges play an
important role in the overall ionization efficiency.

For both C~IV and C~V, the first excited target 
n = 2 threshold(s) lie at a high energy and
the cross sections show  a monotonic decrease over
a relatively large energy range. (Slight oscillations are seen in the ground
level of C V due to some numerical instability; as mentioned earlier,
such oscillations are reduced using larger R-matrix basis set.) The 
resonances at high energies belong to Rydberg series of $n=2,3$ levels.

\section{Recombination cross sections and rate coefficients}

Figs. 3(a,b) present the total recombination cross sections, 
$\sigma_{RC}$, for C~IV and C~V. In contrast to the earlier presentation
for the small energy range (Zhang \etal 1999) to compare with
experiments, the figures display the cross sections from threshold
up to the energy of the highest target threshold, $3d$,
considered in the present work. The resonances belong to different 
n-complexes; these are in close agreeement with experimental data 
(Zhang \etal 1999).

Fig. 4 presents total unified recombination rate coefficients for 
$e~+~C~V \rightarrow C~IV$. The solid curve is the present $\alpha_R$
in relativistic BP approximation, and short-long dashed is earlier
unified rates in LS coupling and where radiation damping effect was not
included (Nahar \& Pradhan 1997). In the high temperature region, the
earlier LS rates significantly overestimat the recombination rate. 
We compare the present BPRM
rates with several other available sets of data, e.g. `experimentally 
derived DR' rates (dot-long dashed curve, Savin 1999; which in fact 
include both the RR+DR contribution - see below), and previous 
theoretical DR rate coefficients in LS coupling 
(dot-dashed curve, Badnell \etal 1990).

Zhang et al (1999) compared in detail the BPRM cross sections with 
experimental data from ion storage rings for $e~+~C~V \rightarrow 
C~IV$, with close agreement in the entire range of measurements for 
both the background (non-resonant) cross sections and resonances. The 
reported experimental data is primarily in the region of low-energy 
resonances that dominate recombination (mainly DR) with H- and He-like 
ions. The recombination rate coefficients, $\alpha_R$, obtained
using the cross sections calculated by Zhang et al. (dotted curve)
agree closely with those of Savin (1999) (dot-long dash curve) who 
used the experimental cross sections to obtain 'experimentally 
derived DR rates'. However, these rates do not include
contributions from  much of the low energy non-resonant RR and very high
energy regions. The total unified $\alpha_R(T)$ (solid curve) which 
include all possible contributions is, therefore, somewhat higher 
than that obtained from limited energy range. 
The LS coupling DR rate by Badnell et al. (1990) (dot-dash curve) is lower than
the others. The dashed and the long-dashed curves in the figure 
are RR rates by Aldrovandi \& Pequignot (1973), and Verner and Ferland 
(1996); the latter agrees with the present rates at lower temperatures.

In Figs. 5 and 6 we show the level-specific recombination 
rate  coefficients into the lowest, and the excited, bound levels of C~IV,
for the $1s^2 \ ns, \ np $ Rydberg series up to n = 10. These are
the first such calculations; level-specific data have been obtained
for all $\ell \leq 9$ and associated $J\pi$ symmetries. 
The behavior of the level-specific rates
mimics that of the total (this is only true for the simple systems 
under consideration; in general the level specifc rates show
significantly different structure for complex ions, as seen in our
previous works). The only distinguishing feature is the DR bump. Since
the numerical computations are enormously involved, particularly related
to the resolution of resonances, 
the absence of any unidentified features in the level-specific rates is
re-assuring.

Total $\alpha_R(T)$ for $e + C VI \rightarrow C V$ are given in 
Table 2, and are plotted in Fig. 7. The total unified recombination 
rate coefficients in the present BPRM calculations are plotted in the
solid curve. The short-long dashed curve represents earlier rates
obtained in LS coupling and with no inclusion of radiation damping 
of low-n autoionizing resonanes (Nahar \& Pradhan 1997). The earlier
LS rates overestimate the recombination rates at high temperatures. 
The dotted curve shows the rate coefficient computed using the Zhang
\etal cross sections in a limited energy range with resonances, i.e.
mainly DR. The DR rate by Shull \& Steenberg (1982; dot-dash curve) 
agrees closely with the dotted curve.
 The dashed and the long-dashed curves 
are RR rates by Aldrovandi \& Pequignot (1973), and Verner and Ferland
(1996); they agrees with the present rates at lower temperatures.

Table 2 presents the unified total BPRM recombination rate coefficients of 
C~IV  and C~V averaged over a Maxwellian distribution.

\subsubsection{Level-specific recombination rate coefficients}

Fig. 8 shows the level-specific rate coefficients for the ground and
the excited n  = 2 levels that are of considerable importance in X-ray
spectroscopy, as they are responsible for the formation of the
w,x,y,z lines from the 4 transitions $1s^2 \ (^1S_0) 
\longleftarrow 1s2p (^1P^o_1), 1s2p (^3P^o_2), 1s2p (^3P^o_1), 1s2s
(^3S_1)$. The present work is particularly relevant to the formation of
these X-ray lines since recombination-cascades from excited levels play
an important role in determining the intensity ratios in 
coronal equilibrium and non-equilibrium plasmas (Pradhan 1985). 

The rates in Fig. 8 differ considerably from those
by Mewe \& Schrijver (1978, hereafter MS) that have been widely 
employed in the calculation
of X-ray spectra of He-like ions (e.g. Pradhan 1982).  We compare with
the direct (RR + DR) rates separately calculated by MS using 
approximate Z-scaled RR and DR rates for
the individual n = 2 levels of He-like ions. 
Their RR rates were from Z-scaled recombinaton rate of He$^+$ given 
by Burgess \& Seaton (1960); the LS coupling data were divided according
to the statistical weights of the fine structure levels. 
Their DR rates
were obtained using averaged autoionization probabilities,
Z-scaled from iron (Z = 26) and calculated with hydrogenic
wavefunctions, together with radiative decay probabilities of
the resonant $2s2p, 2p^2, (2p \ 3s, \ 2p3p, \ 2p3d)$ levels, 
decaying to the final n = 2 levels of He-like ions. The present 
work on the other hand includes DR contributions from all resonances 
up to $2p \ n \ell;  \ n \leq 10, \ell \leq
n-1 $. Figs. 3 and 4 of Zhang \etal (1999) show the detailed
photorecombination cross sections for these resonance complexes.
But the present rate coefficients are much lower (Fig. 8). It is 
surprising that the MS rates are much higher than the
present ones. If we consider, for example, the level-specific rate for
the $1s2s \ (^3S_1)$ level, the MS value includes contributions from
only $2s2p, 3s2p , 3p2p$ autoionizing levels. That the MS values
overestimate the rates is also indicated by the fact that, at Log T =
6.4 (the DR-peak temperature, Fig. 7), the sum of their individual n = 2 
level-specific rates is 1.5 $\times
10^{-12}$ cm$^{-3}$ sec$^{-1}$, compared to our unified {\it total} $\alpha_R$ =
2.27 $\times 10^{-12}$ cm$^{-3}$ sec$^{-1}$ (Table 2; C~V). That would
imply that the MS rates for the n = 2 levels alone account for 2/3
of the total recombination (RR + DR) rate for C~V; which is unlikely.

Fig. 9 presents level specific recombination rate coefficients of $1sns
(^3S)$ Rydberg series of C V levels up to $n$ = 10. The features are
similar to those of C IV. Although resolution
of resonances in each cross section is very cumbersome, the sum of the
level-specific rate coefficients, together with the DR contribution,
agrees within a few percent of the total recombination rate
coefficient.

 Recombination-cascade matrices may now be constructed for C~IV and C~V, 
and effective recombination rates into specific levels obtained
accurately, using the direct recombination rates into levels with $n \leq
10, \ell \leq n-1 $ (Pradhan 1985). The present data is more than sufficient
for extrapolation to high-n,$\ell$ necessary to account for 
cascade contributions. Also needed are the radiative transition 
probabilities for all
fine structure levels of C~IV and C~V, up to the n = 10 levels. 
They have also been calculated using the
BPRM method, and will be available shortly (Nahar, in preparation). These
data will be similar to that for Fe~XXIV and Fe~XXV calculated earlier under
the Iron Project  (Nahar and Pradhan 1999).

We discuss below some of the important atomic effects relavant to
the present calculations in particular, and electron-ion recombination
in general.

\subsection{Resolution and radiation damping of resonances}

  It is important to resolve near-threshold resonances ($\nu \leq 10$)
at an adequately
fine energy mesh in order to (a) compute accurately their contribution to
the rate coefficient or the averaged cross section, and (b) to determine
the radiative and autoionization rates through the fitting procedure
referred to eariler
(Sakimoto et al. 1990, Pradhan and Zhang 1997, Zhang \etal 1999). Resonances
that are narrower than the energy intervals chosen have
 very low autoionization rates and are mostly damped out; their 
contribution to (a) should be small.

\subsection{Intereference between resonant (DR) and non-resonant
(RR) recombination}

 In general there is quantum mechanical interference between the
resonant and the non-resonant components of the wavefunction expansion.
Close coupling photoioniation calculations for strongly coupled 
near-neutral atomic systems cross sections considerable overlap between
members of several Rydberg series of resonances that coverge on to the
excited, coupled target levels.

 Unified (e~+~ion) rates have been calculated for over 40 atoms and ions
of astrophysical interest. The resonance structures 
in a number of these are extremely complicated and show considerable 
interference. 
The large number of  photoionization calculations
under the Opacity Project, for the ground state and the excited states
(typically, a few hundred excited states for each atom or ion),
show overlapping resonances to often dominate the cross sections. These
are archived in the Opacity Project database TOPBASE and may be accessed 
on-line via the Website: http://heasarc.gsfc.nasa.gov, or via the link from
www.astronomy.ohio-state.edu/$\sim$pradhan). 
Furthermore,
the excited state
cross sections of excited metastable states may exhibit even more
extensive resonance structures than the ground state (e.g.
in photoionization of O~III; Luo \etal 1989). Strictly speaking, one
needs to consider the {\it partial} photoionization cross sections for
each of the three cases (dipole photoionization): $\Delta S = 0, \Delta
l = 0, \pm 1$ in LS coupling, or $\Delta J = 0, \pm 1$ in BPRM calculations. 
Although these are not often tabulated or displayed, an
examination thereof reveals that for near-neutrals each partial cross
section shows overlapping resonances. As the ion charge
increases, the resonances separate out, the intereference decreases, and 
isolated resonance approximations may be used (Pindzola \etal 1992, Zhang 
1998).

Therefore, in general, we expect interference between the non-resonant 
RR and resonant DR recombination processes, particularly for
many-electron systems where a separation between the two processes is
unphysical and imprecise. Of course experimentally such a division
is artificial and is not possible. For example, the recent experimental
measurements on electron recombination with Fe~XVIII to Fe~XVII (Savin
\etal 1999) clearly show the near threshold cross section to be dominated
by the non-resonant recombination (RR) towards E $\rightarrow$ 0, with
superimposed resonance structures (the DR contribution) at higher
energies. 
In a recent work Savin (1999) cited the 
work of Pindzola \etal (1992) to state that the
effect of interferences is small (Pindzola \etal did not
carry out close coupling calculations such as in the Opacity Project).
 Although this is not true in general, 
for highly ionized few-electron ions one expects sufficient resonance
separation so that an independent treatment of RR and DR may be
accurate; such is the case for recombination with H-like and He-like
ions. 
The present 
unified treatment accounts for interference effects
in an ab initio manner, and (e~+~ion) rates have been calculated
for over 40 atoms and ions of astrophysical interest (heretofore, in LS
coupling).

\subsection{Comparision with experimental data and uncertainties}

 Although experimental results are available for relatively few ions in
limited energy ranges, and
mostly for simple atomic systems such as the H-like and He-like
ions, they are very useful for the calibration of theoretical cross
sections. There is very good agreement with experimentally measured cross 
sections for
electron recombination to C~IV, C~V, and O~VII, as discussed in detail
by Zhang \etal (1999). However, the energy range of experimental measurements
is much smaller; the theoretical calculations are from
E = 0 to very high energies necessary to obtain rate coefficients up to
T = $10^9$K. For recombination with H- and
He-like ions most (but not all) of the rates depends on relatively few
low-n complex of resonances in the low-energy region covered by
experiments. In a recent work Savin (1999) used the experimental cross
sections for recombination with C~V to C~IV, and O~VIII to O~VII,  
to obtain `experimentally derived DR' rate coeffcients and
compared those with several sets of theoretical data including those in
papers I and II. (Strictly speaking these rates include `RR + DR'
contributions since the experimental cross sections
always include both, and which the unified
method aims to obtain). Whereas the previous LS coupling results for C~IV
(paper I) were 43\% higher than Savin's values, the results for O~VII
were in reasonable agreement,  $\sim$20\% higher, within estimated
uncertainties.

The agreement between the unified rates and the experimentally derived
DR rates is within our 10-20\% in the region (i.e. at temperatures)
where DR contribution peaks, around Log(T) = 6.3 for C~IV (Fig. 2).
As the reported experimental data did not extend to low energies, where
the non-resonant RR contribution dominates, the unified rates are 
higher towards lower temperatures from
the DR peak, and deviate in a predictably straightforward manner from the 
DR-only results. The `experimentailly derived' data by Savin (1999) was
obtained with a limited energy range, while our present results include
a much larger range. Therefore our results are higher. The dotted curve
in Fig. 4, which we also obtained over a limited energy range, agrees
well with the dot-long-dashed curve (Savin 1999).
 The differences are within our estimate of uncertainty in the
present results, up to 20\%. Given that the experimental cross sections
are also likely to be uncertain to about this range, the agreement seems
remarkably good.

While the present unified cross sections can be compared directly
with experimental measurements, and the new rate coefficients are in
good agreement with the experimentaly derived DR rates for recombination
with simple ions such as the H- and He-like, the experimental
data may represent a lower bound
on the field-free theoretical recombination rates owing to (a)
high-n and $\ell$ ionization reducing the 'DR' peak,
and (b) limited energy range in experiments.

\subsection{General features of (e + ion) recombination rates}

The non-resonant `RR' recombination peaks as E, T $\rightarrow$ 0.
This is due to the dominant contribution from an infinite number of high
Rydberg states of the (e~+~ion) system into which the slow moving electron 
may recombine. At low-E and T, the total $log_{10}(\alpha_R(T)$) is shown 
as a straight line on the Log-Log scale due to the exponential Maxwellian 
damping factor exp(-E/kT). It is not entirely trivial to compute the 
low-E and T contributions (that we refer to as ``high-n top-up").
We adapt the accurate numerical procedure developed by Storey \&
Hummer (1992) to calculate the n,$\ell$ 
hydrogenic photoionization cross section for $11 \leq n \leq \infty$ 
(Nahar 1996).
It is noted that the high-n top-up also represents the otherwise missing
background contribution due to high-n resonant recombination (DR). Although
this background contribution is small (negligibly so for the H- and
He-like ions), it is included in the unified treatment. 

 The resonant contribution (DR) peaks at higher E and T corresponding to
the excitation energies and temperatures of the strong dipole transition(s) in
the core ion. This is the broad peak in $\alpha_R(T)$.

\subsection{Ionization fractions of Carbon}

 Fig. 10 presents coronal ionization fractions of C using the new BPRM
recombinaton rates for C~IV and C~V (solid lines). Also given are the
results (dashed lines) from Arnaud and Rothenflug (1985), and previous
results (dotted lines) using LS coupling rates from (Nahar and Pradhan 1997).
Differences with both sets of data may be noted for C~IV, C~V, C~VI, and
C~VII. The most significant change (enhancement) is for C~VI, 
owing to the decrease in C~V recombination rate, and the new ionization 
fractions appear to be in better agreement with Arnaud and Rothenflug (1985)
 than the Nahar and Pradhan (1997) results.  Also discernible is the
steeper fall-off in the C~V ionization fraction on the high temperatures
side.

\section{CONCLUSION}

 New relativistic calculations are presented for the total, unified
(e~+~ion) rates coefficients for C~IV and C~V of interest in X-ray astronomy. As
the photo-recombination cross sections in the dominant low-energy region
have earlier been shown to be in very good agreeement with experiments
(Zhang \etal 1999), it is expected that the present rates should be
definitive, with an uncertainty that should not exceed 10--20\%.

The unified theoretical formulation
and experimental measurements both
suggest that the unphysical and imprecise division of the
recombination process into 'radiative recombination (RR)' and
'di-electronic recombination (DR)' be replaced by
'non-resonant' and 'resonant' recombination, since these are naturally
inseparable.

 Further calculations are in progress for Oxygen (O~VI and O~VII) and
Iron (Fe~XXIV and FeXXV).

The available data includes: 

(A) Photoionization cross sections for bound fine structure levels
of C~IV and C~V up to n = 10 -- both total and partial (into the ground
level of the residual ion). !!!!

(B) Total, unified recombination rates for C~IV and C~V, and level-specific 
recombination rate coefficients for levels up to n = 10. 

All photoionization and recombination data are available electronically
from the first author at: nahar@astronomy.ohio-state.edu. The total
recombination rate coefficients are also available from the Ohio State
Atomic Astrophysics website at: www.astronomy.ohio-state.edu/$\sim$pradhan.

%

\acknowledgments

This work was supported partially by grants from NSF (AST-9870089)
 and NASA (NAG5-8423). 
The computational work was carried out on the
Cray T94 at the Ohio Supercomputer Center.

\clearpage

\begin{table}
\caption{Target terms in the eigenfunction expansions of C V and C VI.
The target energies are in eV.
}
\scriptsize
\begin{tabular}{llll}
\hline
\multicolumn{2}{c}{C V} & \multicolumn{2}{c}{C VI} \\
\hline
1s$^2(^1{\rm S}_0)$     & 0.0    &
1s$(^2{\rm S}_{1/2})$   &   0.00 \\
1s2s$(^3{\rm S}_1)$     & 298.73 &
2s$(^2{\rm S}_{1/2})$   & 367.36  \\
1s2s$(^1{\rm S}_0)$     & 304.38 &
2p$(^2{\rm P}\o_{1/2})$ & 367.36  \\
1s2p$(^3{\rm P}\o_0)$   & 304.39 &
2p$(^2{\rm P}\o_{3/2})$ & 367.42 \\
1s2p$(^3{\rm P}\o_1)$   & 304.39 &
3s$(^2{\rm S}_{1/2})$   & 435.41 \\
1s2p$(^3{\rm P}\o_2)$   & 304.41 &
3p$(^2{\rm P}\o_{1/2})$ & 435.41 \\
1s2p$(^1{\rm P}\o_1)$   & 307.90 &
3p$(^2{\rm P}\o_{3/2})$ & 435.43 \\
1s3s$(^3{\rm S}_1)$     & 352.05 &
3d$(^2{\rm D}_{3/2})$   & 435.43 \\
1s3s$(^1{\rm S}_0)$     & 353.49 &
3d$(^2{\rm D}_{5/2})$   & 435.43 \\
1s3p$(^3{\rm P}\o_0)$   & 353.52 & & \\
1s3p$(^3{\rm P}\o_1)$   & 353.52 & & \\
1s3p$(^3{\rm P}\o_2)$   & 353.52 & & \\
1s3p$(^1{\rm P}\o_1)$   & 354.51 & & \\
\multicolumn{2}{c}{13-CC}  & \multicolumn{2}{c}{9-CC} \\
\hline
\end{tabular}
\end{table}

\pagebreak

\begin{table}
\caption{Total recombination rate coefficients, $\alpha_R(T)$, of C IV
and C V. }
\scriptsize
\begin{tabular}{crrcrr}
\hline
$log_{10}T$ & \multicolumn{2}{c}{$\alpha_R(cm^3s^{-1})$} &
$log_{10}T$ & \multicolumn{2}{c}{$\alpha_R(cm^3s^{-1})$}\\
(K) & \multicolumn{1}{c}{C IV} & \multicolumn{1}{c}{C V} &
(K) & \multicolumn{1}{c}{C IV} & \multicolumn{1}{c}{C V} \\
\hline
  1.00 &  5.86E-10 & 9.87E-10 &
  5.10 &  1.32E-12 & 2.84E-12 \\
  1.10 &  5.14E-10 & 8.68E-10 &
  5.20 &  1.10E-12 & 2.42E-12 \\
  1.20 &  4.51E-10 & 7.62E-10 &
  5.30 &  9.17E-13 & 2.06E-12 \\
  1.30 &  3.95E-10 & 6.69E-10 &
  5.40 &  7.61E-13 & 1.75E-12 \\
  1.40 &  3.45E-10 & 5.88E-10 &
  5.50 &  6.34E-13 & 1.49E-12 \\
  1.50 &  3.02E-10 & 5.15E-10 &
  5.60 &  5.42E-13 & 1.27E-12 \\
  1.60 &  2.63E-10 & 4.51E-10 &
  5.70 &  5.11E-13 & 1.11E-12 \\
  1.70 &  2.30E-10 & 3.94E-10 &
  5.80 &  5.76E-13 & 1.04E-12 \\
  1.80 &  2.01E-10 & 3.45E-10 &
  5.90 &  7.53E-13 & 1.11E-12 \\
  1.90 &  1.75E-10 & 3.01E-10 &
  6.00 &  1.01E-12 & 1.33E-12 \\
  2.00 &  1.52E-10 & 2.63E-10 &
  6.10 &  1.28E-12 & 1.65E-12 \\
  2.10 &  1.32E-10 & 2.29E-10 &
  6.20 &  1.48E-12 & 1.97E-12 \\
  2.20 &  1.15E-10 & 2.00E-10 &
  6.30 &  1.57E-12 & 2.19E-12 \\
  2.30 &  1.00E-10 & 1.74E-10 &
  6.40 &  1.54E-12 & 2.27E-12 \\
  2.40 &  8.68E-11 & 1.52E-10 &
  6.50 &  1.42E-12 & 2.19E-12 \\
  2.50 &  7.55E-11 & 1.32E-10 &
  6.60 &  1.25E-12 & 2.00E-12 \\
  2.60 &  6.53E-11 & 1.15E-10 &
  6.70 &  1.05E-12 & 1.74E-12 \\
  2.70 &  5.66E-11 & 9.99E-11 &
  6.80 &  8.48E-13 & 1.45E-12 \\
  2.80 &  4.90E-11 & 8.67E-11 &
  6.90 &  6.69E-13 & 1.18E-12 \\
  2.90 &  4.24E-11 & 7.52E-11 &
  7.00 &  5.17E-13 & 9.29E-13 \\
  3.00 &  3.66E-11 & 6.52E-11 &
  7.10 &  3.93E-13 & 7.20E-13 \\
  3.10 &  3.16E-11 & 5.65E-11 & 
  7.20 &  2.94E-13 & 5.49E-13 \\
  3.20 &  2.73E-11 & 4.90E-11 &
  7.30 &  2.18E-13 & 4.13E-13 \\
  3.30 &  2.35E-11 & 4.24E-11 &
  7.40 &  1.60E-13 & 3.08E-13 \\
  3.40 &  2.02E-11 & 3.67E-11 &
  7.50 &  1.17E-13 & 2.28E-13 \\
  3.50 &  1.74E-11 & 3.17E-11 &
  7.60 &  8.47E-14 & 1.67E-13 \\
  3.60 &  1.50E-11 & 2.74E-11 &
  7.70 &  6.12E-14 & 1.23E-13 \\
  3.70 &  1.29E-11 & 2.37E-11 &
  7.80 &  4.41E-14 & 8.93E-14 \\
  3.80 &  1.10E-11 & 2.04E-11 &
  7.90 &  3.17E-14 & 6.49E-14 \\
  3.90 &  9.46E-12 & 1.76E-11 &
  8.00 &  2.27E-14 & 4.70E-14 \\
  4.00 &  8.10E-12 & 1.52E-11 &
  8.10 &  1.62E-14 & 3.40E-14 \\
  4.10 &  6.93E-12 & 1.31E-11 &
  8.20 &  1.16E-14 & 2.46E-14 \\
  4.20 &  5.92E-12 & 1.13E-11 &
  8.30 &  8.27E-15 & 1.77E-14 \\
  4.30 &  5.05E-12 & 9.69E-12 &
  8.40 &  5.90E-15 & 1.28E-14 \\
  4.40 &  4.30E-12 & 8.33E-12 &
  8.50 &  4.20E-15 & 9.20E-15 \\
  4.50 &  3.66E-12 & 7.16E-12 &
  8.60 &  2.99E-15 & 6.62E-15 \\
  4.60 &  3.10E-12 & 6.15E-12 &
  8.70 &  2.13E-15 & 4.76E-15 \\
  4.70 &  2.63E-12 & 5.28E-12 &
  8.80 &  1.52E-15 & 3.42E-15 \\
  4.80 &  2.22E-12 & 4.53E-12 &
  8.90 &  1.08E-15 & 2.46E-15 \\
  4.90 &  1.87E-12 & 3.88E-12 &
  9.00 &  7.67E-16 & 1.77E-15 \\
  5.00 &  1.57E-12 & 3.32E-12 &
  & & \\
\hline
\end{tabular}
\end{table}


\clearpage

%
%

\def\amp{{Adv. At. Molec. Phys.}\ }
\def\apj{{ Astrophys. J.}\ }
\def\apjs{{Astrophys. J. Suppl. Ser.}\ }
\def\apjl{{Astrophys. J. (Letters)}\ }
\def\aj{{Astron. J.}\ }
\def\aa{{Astron. Astrophys.}\ }
\def\aasup{{Astron. Astrophys. Suppl.}\ }
\def\adndt{{At. Data Nucl. Data Tables}\ }
\def\cpc{{Comput. Phys. Commun.}\ }
\def\jqsrt{{J. Quant. Spectrosc. Radiat. Transfer}\ }
\def\jpb{{Journal Of Physics B}\ }
\def\pasp{{Pub. Astron. Soc. Pacific}\ }
\def\mn{{Mon. Not. R. astr. Soc.}\ }
\def\pra{{Physical Review A}\ }
\def\prl{{Physical Review Letters}\ }
\def\zpds{{Z. Phys. D Suppl.}\ }
\def\adndt{Atomic Data And Nuclear Data Tables}

%

\newpage

\begin{figure}
\centering
\psfig{figure=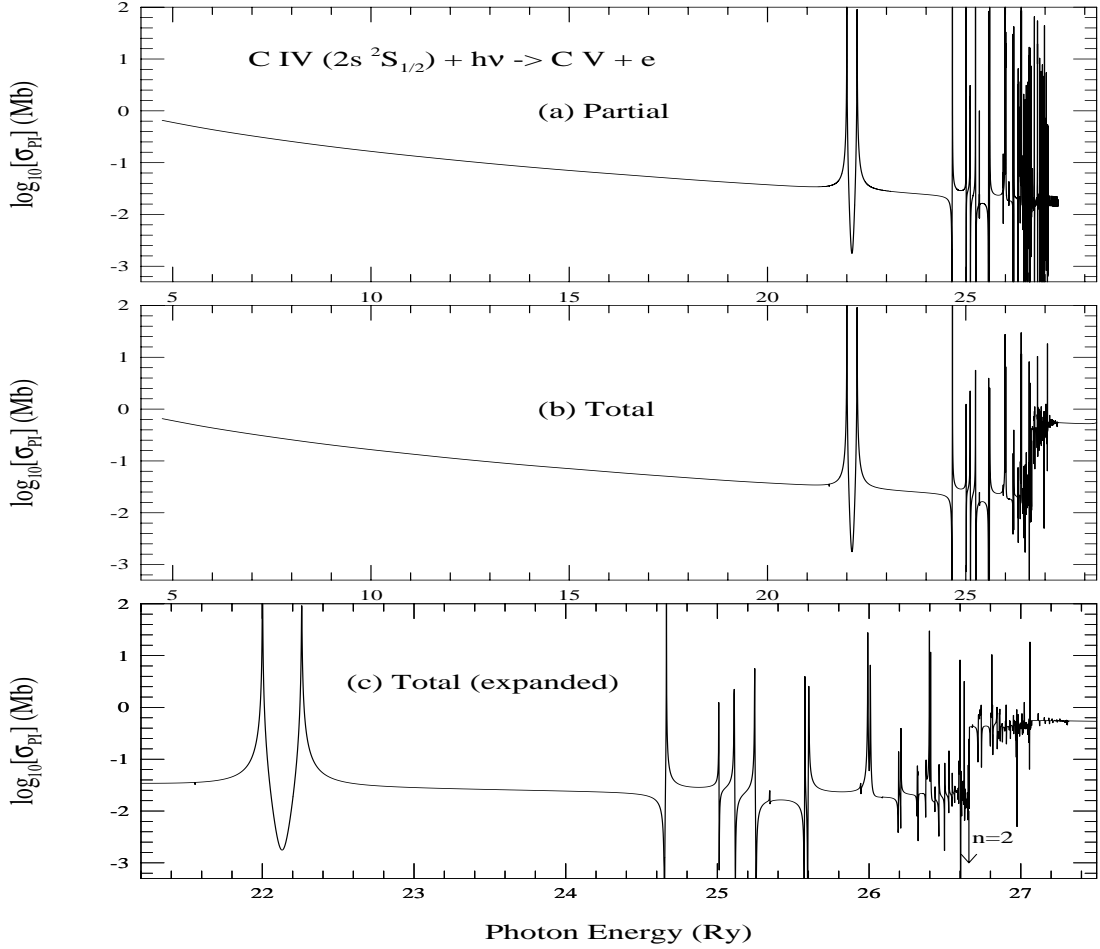,height=15.0cm,width=18.0cm}
\caption{Photoionization of the
ground state $1s^2 \ 2s \ (^2S_{1/2})$ of C~IV: partial cross section
into the ground level $1s^2 \ (^1S_0)$ of C~V (b); total cross section (a);
an expanded view of the resonances and inner-shell thresholds (c).
The large jump in (a) corresponds to the K-shell ionization edge. (The
total cross sections are at a coarser mesh than the partial ones,
therefore some of the resonance heights are smaller).)}
\end{figure}

\begin{figure}
\centering
\psfig{figure=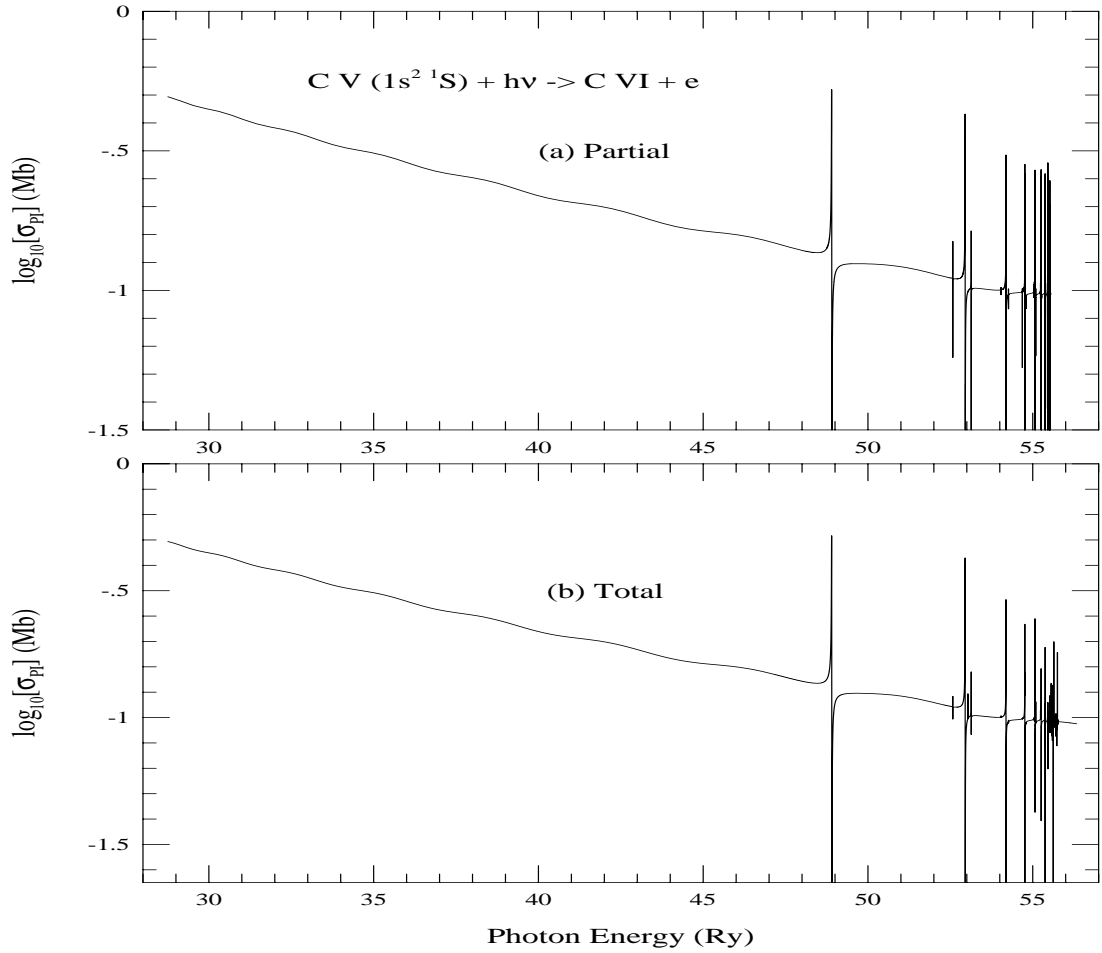,height=15.0cm,width=18.0cm}
\caption{Photoionization of the
ground state $1s^2 \ (^1S_0)$ of C~V: partial cross section into the
ground level $1s \ (^2S_{1/2})$ of C~VI (a); total cross section (b).
(The total cross sections are at a coarser mesh than the partial ones,
therefore some of the resonance heights are smaller.)}
\end{figure}

\begin{figure}
\centering
\psfig{figure=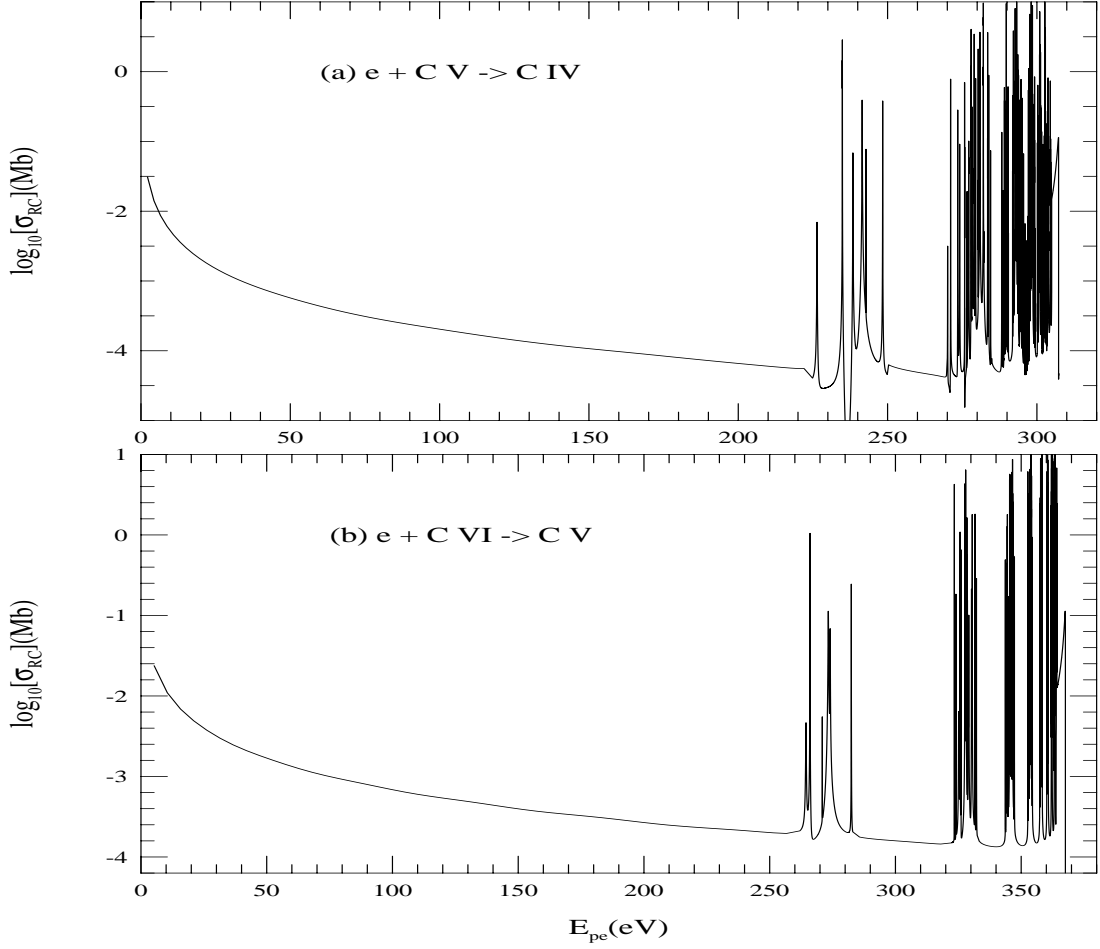,height=15.0cm,width=18.0cm}
\caption{Total unified (e + ion) photo-recombination cross 
sections, $\sigma_{RC}$, of (a) C~IV and (b) C~V. Note that the $\sigma_{RC}$
exhibit more resonance structures than the corresponding ground level
$\sigma_{PI}$ in Figs. 1 and 2, since the former are summed over the
ground {\it and} many excited recombined levels.}
\end{figure}

\begin{figure}
\centering
\psfig{figure=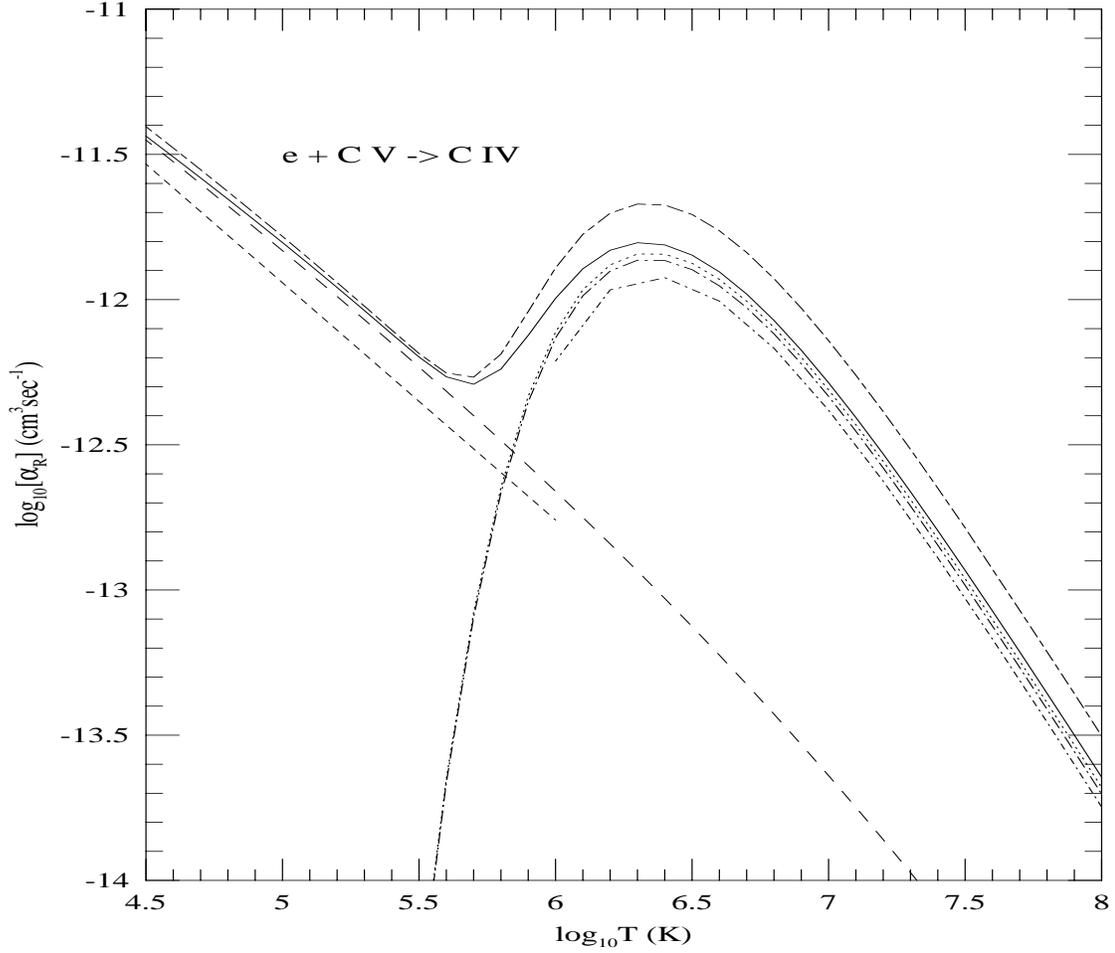,height=15.0cm,width=18.0cm}
\caption{ (e + C V) $\longrightarrow$ C~IV; total unified rate coefficients:
BPRM (with fine structure), solid curve; LS coupling, short and long
dashed curve; from Zhang et al (1999), dotted;
from Savin (1999), dot-long dash curve; LS coupling 
DR rates from Badnell et al (1990),
dot-dash curve; RR rates from Aldrovandi and Pequignot (1973),
short-dash; RR rates from Verner and Ferland (1996), long-dash.} 
\end{figure}

\begin{figure}
\centering
\psfig{figure=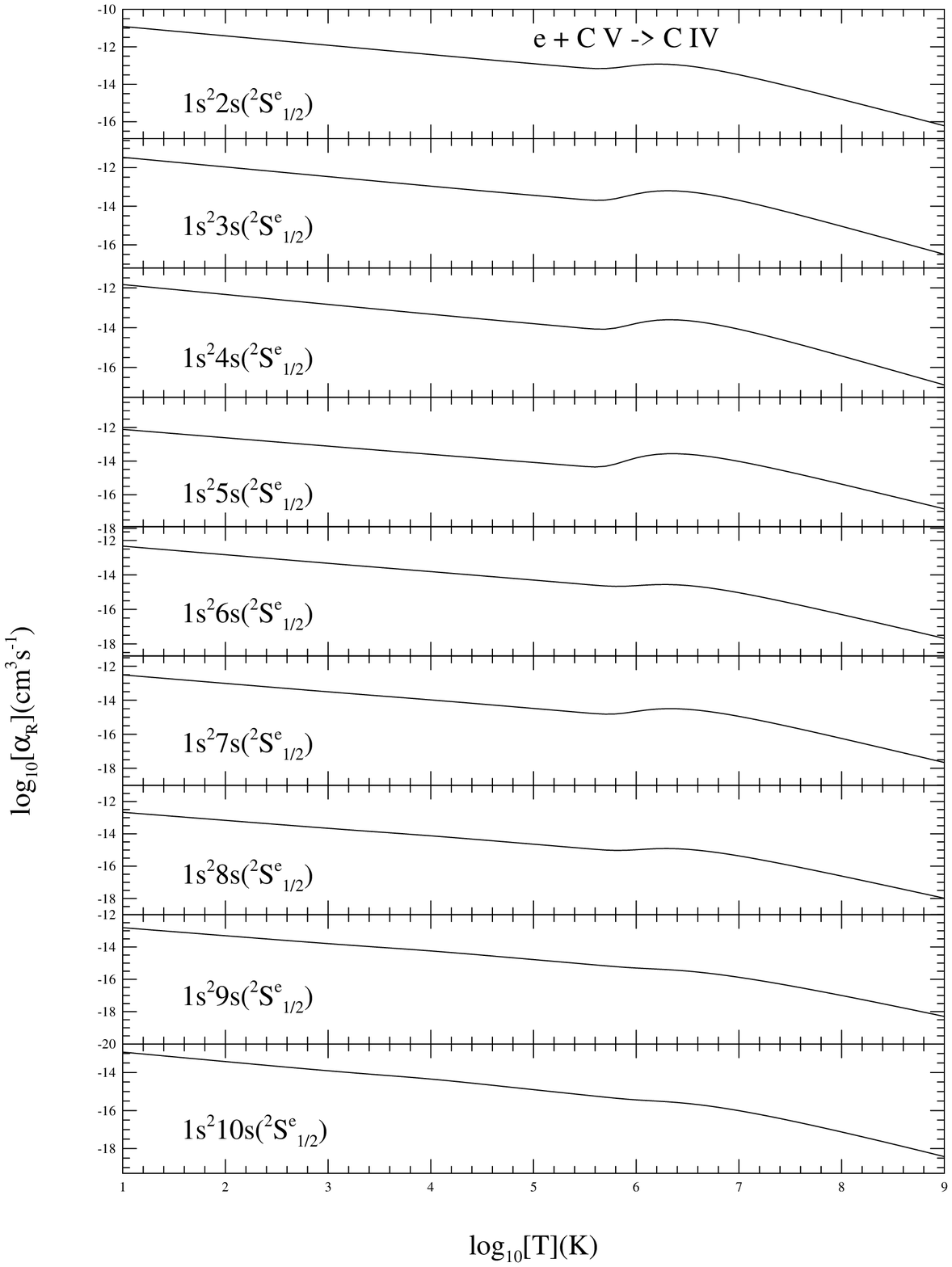,height=15.0cm,width=18.0cm}
\caption{ Level-specific recombination rate coefficients of  e + C~V 
$\longrightarrow$ C~IV into the ground and excited levels of 
the $1s^2 \ ns$ Rydberg series, $n \leq 10$.}
\end{figure}

\begin{figure}
\centering
\psfig{figure=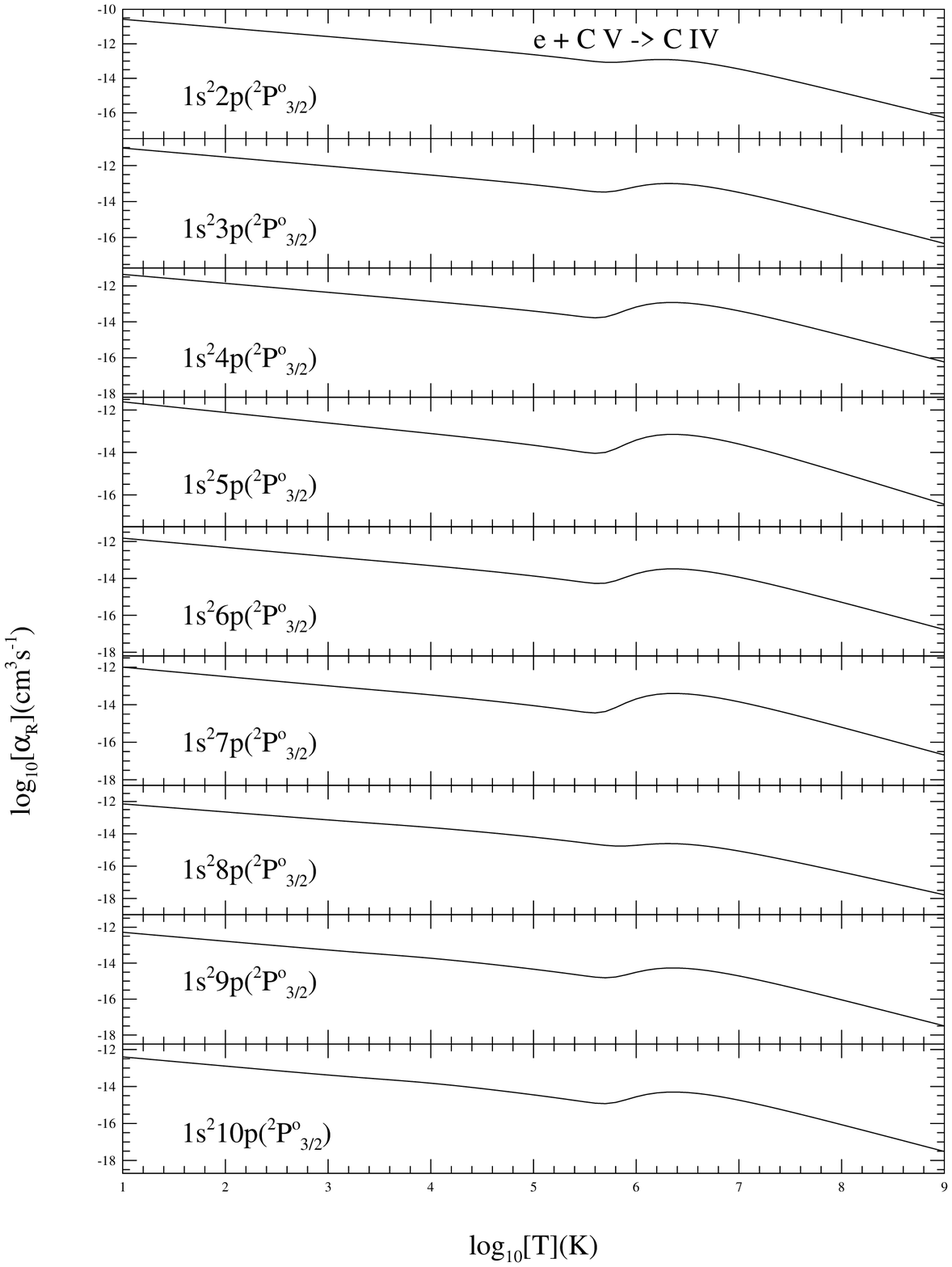,height=15.0cm,width=18.0cm}
\caption{ Level-specific recombination rate coefficients of  e + C~V 
$\longrightarrow$ C~IV into the ground and excited levels of 
the $1s^2 \ np$ Rydberg series, $n \leq 10$.}
\end{figure}

\begin{figure}
\centering
\psfig{figure=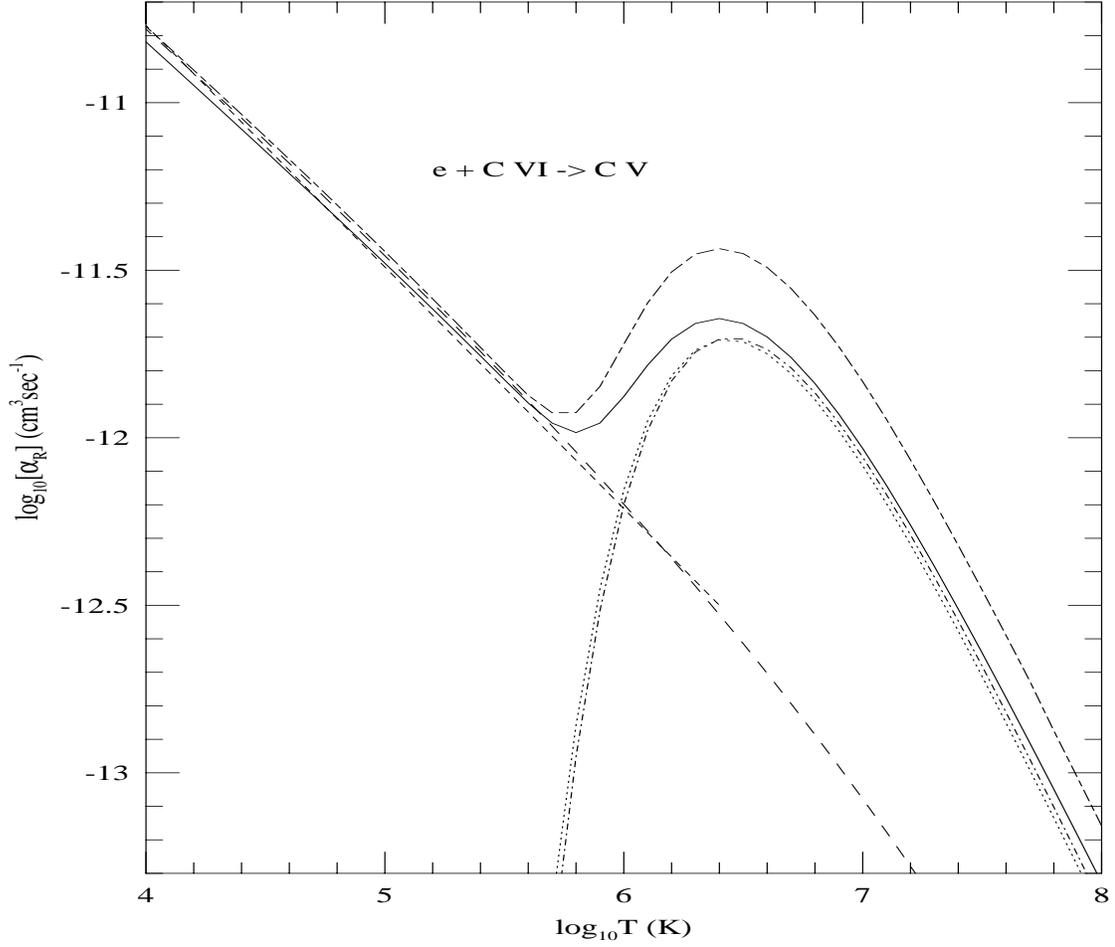,height=15.0cm,width=18.0cm}
\caption{ (e + C VI) $\longrightarrow$ C V; total unified rate coefficients:
BPRM with fine structure - solid, LS coupling - short and long
dashed, using Zhang \etal (1999) cross sections in the same 
range as experimental
data - dotted; DR rates by Shull \& Steenberg (1982) - dot-dash; 
RR rates: Aldrovandi and Pequignot (1973) - short-dash; Verner and 
Ferland (1996) - long-dash.} 
\end{figure}

\begin{figure}
\centering
\psfig{figure=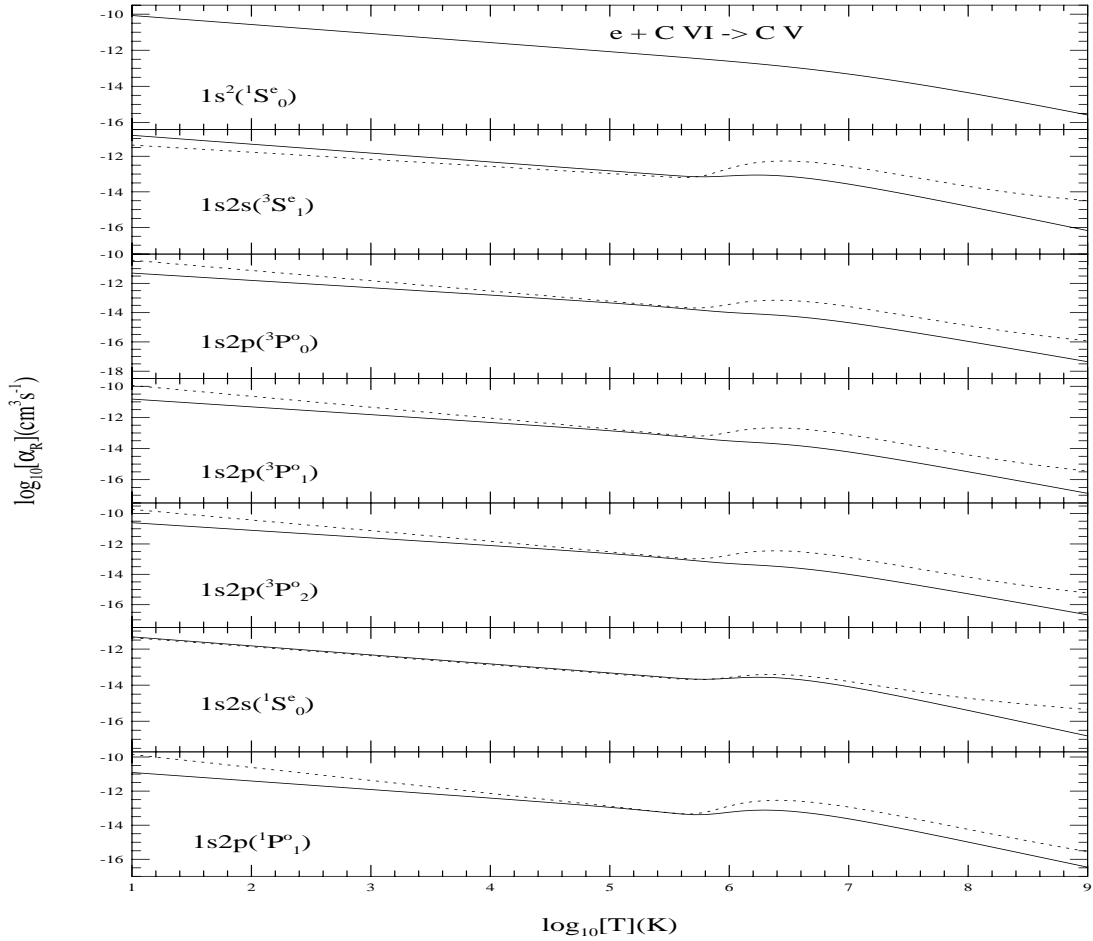,height=15.0cm,width=18.0cm}
\caption{Direct level-specific recombination rate coefficients for  (e + C~VI)
$\rightarrow$ C~V into the ground and the excited  n = 2 levels of 
C~V - solid; Mewe and Schrijver (1978) - dotted.}
\end{figure}

\begin{figure}
\centering
\psfig{figure=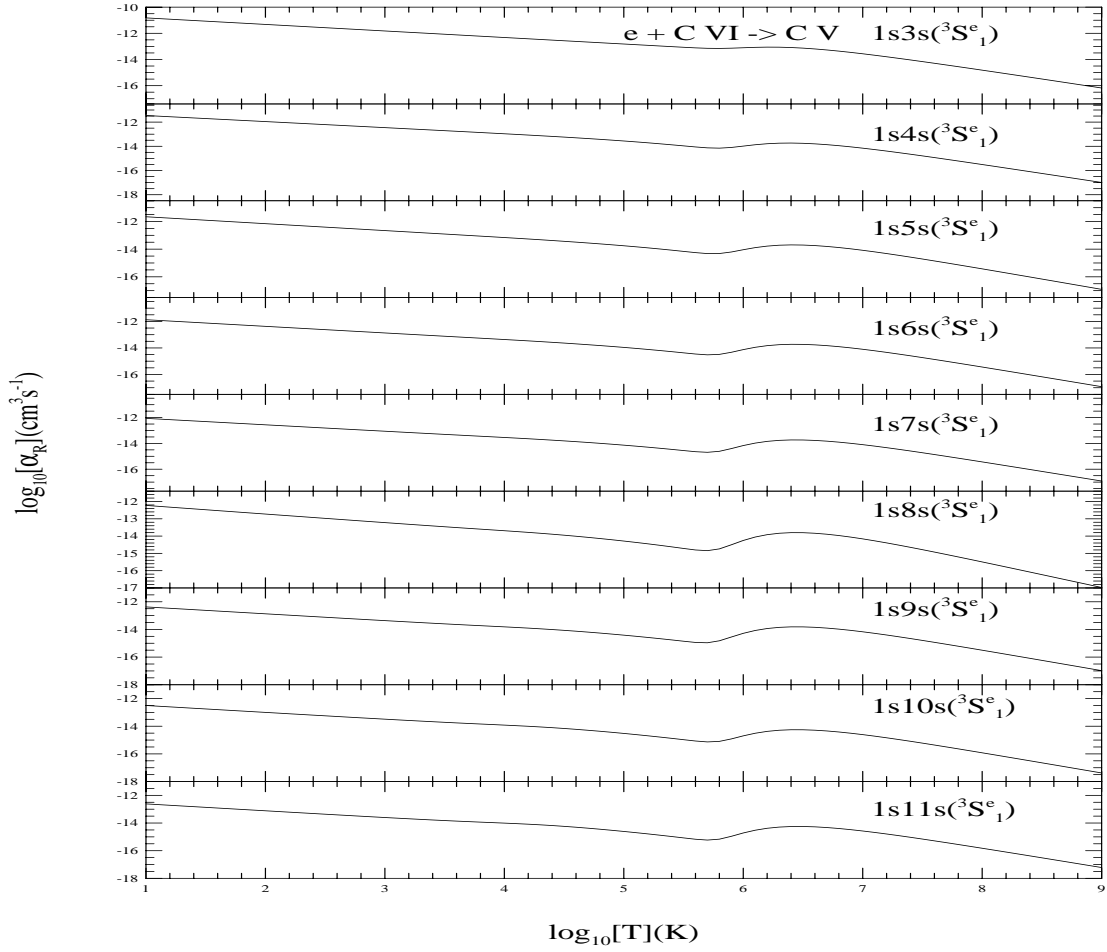,height=15.0cm,width=18.0cm}
\caption{Level-specific recombination rate coefficients for  (e + C~VI)
$\rightarrow$ C~V into the excited levels of 
the $1s \ ns \ (^3S_1)$ Rydberg series, $n \leq 11$.}
\end{figure}

\begin{figure}
\centering
\psfig{figure=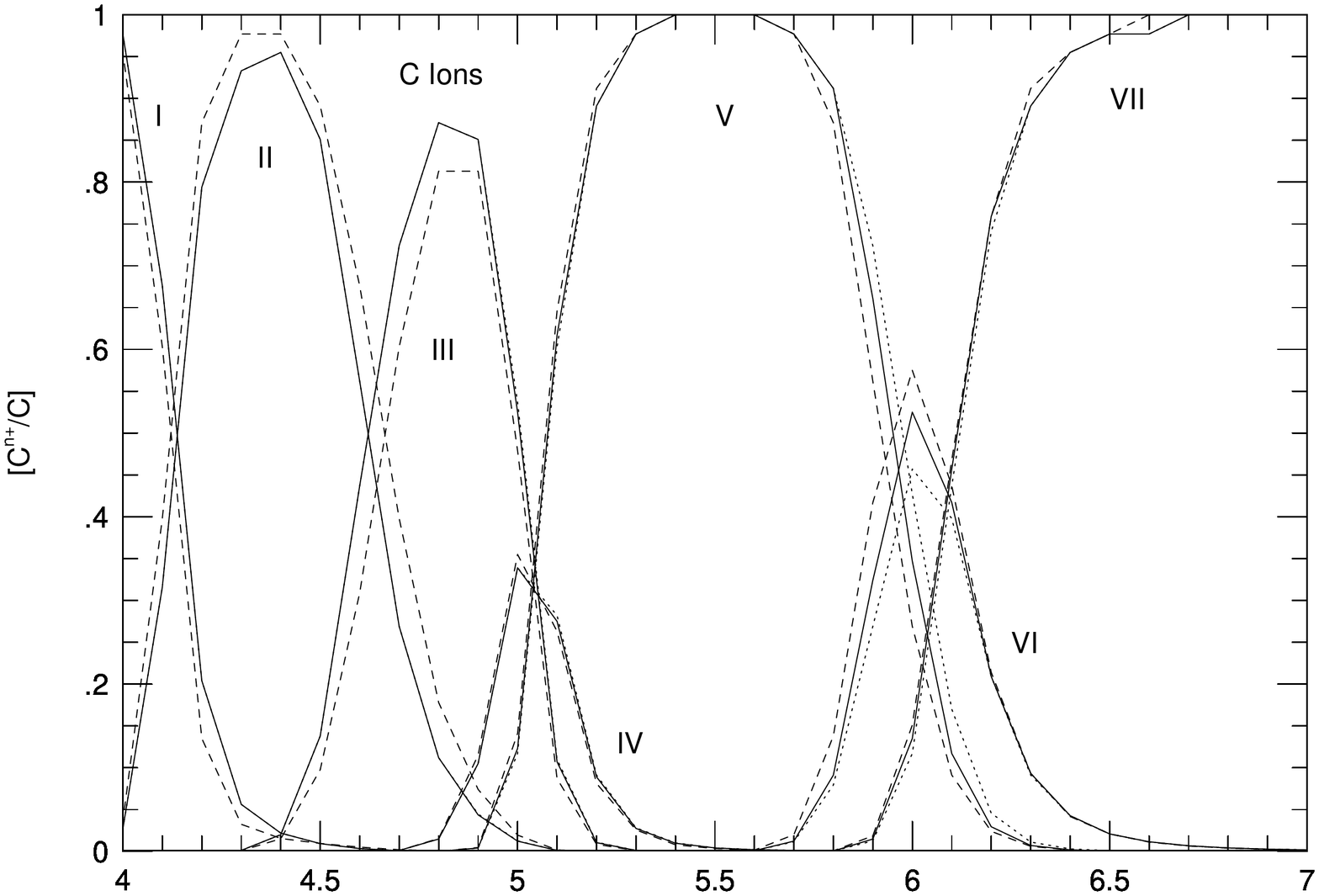,height=15.0cm,width=18.0cm}
\caption{Ionization fractions of carbon in coronal equilibrium using
the present recombination rates for C~IV and C~V -- solid curve; using
the LS coupling rates by Nahar and Pradhan (1997) -- dashed curve;
ionization fractions from Arnaud and Rothenflug (1985) -- dotted
curve.} 
\end{figure}

\end{document}